\documentclass[longauth,traditabstract]{aa}
\usepackage{epsfig,times}
\usepackage{multirow}
\usepackage{graphicx}
\usepackage{natbib}
\usepackage{txfonts}

\def\mincir{\raise -2.truept\hbox{\rlap{\hbox{$\sim$}}\raise5.truept \hbox{$<$}\ }}
\def\mincireq{\hbox{\raise0.5ex\hbox{$<\lower1.06ex\hbox{$\kern-1.07em{\sim}$}$}}}
\def\magcir{\raise-2.truept\hbox{\rlap{\hbox{$\sim$}}\raise5.truept \hbox{$>$}\ }}
\def\gr{\kern 2pt\hbox{}^\circ{\kern -2pt K}} 

\def\_{\thinspace}

\def\be{\begin{equation}}
\def\ee{\end{equation}}

\begin{document}

\title{Simultaneous multi-frequency observation of the unknown redshift blazar PG\,1553+113 in March-April 2008}

\author{J.~Aleksi\'{c}\inst{1} \and
 H.~Anderhub\inst{2} \and
 L.~A.~Antonelli\inst{3} \and
 P.~Antoranz\inst{4} \and
 M.~Backes\inst{5} \and
 C.~Baixeras\inst{6} \and
 S.~Balestra\inst{4} \and
 J.~A.~Barrio\inst{4} \and
 D.~Bastieri\inst{7} \and
 J.~Becerra Gonz\'alez\inst{8}
% ...
\\
(The MAGIC Collaboration)
\\
E.~Pian\inst{33,}\inst{39} \and
V.~Bianchin\inst{30} \and 
F.~D'Ammando\inst{31} \and
G.~Di Cocco\inst{30} \and
D.~Fugazza\inst{32}  \and
G.~Ghisellini\inst{32} \and
O.~M.~Kurtanidze\inst{38} \and
C.~M.~Raiteri\inst{34} \and 
G.~Tosti\inst{35} \and
A.~Treves\inst{36} \and 
S.~Vercellone\inst{37} \and
M.~Villata\inst{34}
\\
}
            
\offprints{N. Mankuzhiyil, \email{nijil.mankuzhiyil@uniud.it}}

\institute{IFAE, Edifici Cn., Campus UAB, E-08193 Bellaterra, Spain
 \and ETH Zurich, CH-8093 Switzerland
 \and INAF National Institute for Astrophysics, I-00136 Rome, Italy
 \and Universidad Complutense, E-28040 Madrid, Spain
 \and Technische Universit\"at Dortmund, D-44221 Dortmund, Germany
 \and Universitat Aut\`onoma de Barcelona, E-08193 Bellaterra, Spain
 \and Universit\`a di Padova and INFN, I-35131 Padova, Italy
 \and Inst. de Astrof\'{\i}sica de Canarias, E-38200 La Laguna, Tenerife, Spain
 \and University of \L\'od\'z, PL-90236 Lodz, Poland
 \and Tuorla Observatory, University of Turku, FI-21500 Piikki\"o, Finland
 \and Deutsches Elektronen-Synchrotron (DESY), D-15738 Zeuthen, Germany
 \and Max-Planck-Institut f\"ur Physik, D-80805 M\"unchen, Germany
 \and Universit\`a  di Siena, and INFN Pisa, I-53100 Siena, Italy
 \and Universitat de Barcelona (ICC/IEEC), E-08028 Barcelona, Spain
 \and Universit\"at W\"urzburg, D-97074 W\"urzburg, Germany
 \and Depto. de Astrofisica, Universidad, E-38206 La Laguna, Tenerife, Spain
 \and Universit\`a di Udine, and INFN Trieste, I-33100 Udine, Italy
 \and Institut de Ci\`encies de l'Espai (IEEC-CSIC), E-08193 Bellaterra, Spain
 \and Inst. de Astrof\'{\i}sica de Andaluc\'{\i}a (CSIC), E-18080 Granada, Spain
 \and Croatian MAGIC Consortium, Institute R. Boskovic, University of Rijeka and University of Split, HR-10000 Zagreb, Croatia
 \and University of California, Davis, CA-95616-8677, USA
 \and Inst. for Nucl. Research and Nucl. Energy, BG-1784 Sofia, Bulgaria
 \and INAF/Osservatorio Astronomico and INFN, I-34143 Trieste, Italy
 \and ICREA, E-08010 Barcelona, Spain
 \and Universit\`a  di Pisa, and INFN Pisa, I-56126 Pisa, Italy
 \and supported by INFN Padova
 \and now at: Centro de Investigaciones Energ\'eticas, Medioambientales y Tecnol\'ogicas (CIEMAT), Madrid, Spain
 \and now at: Max-Planck-Institut f\"ur Kernphysik, D-69029 Heidelberg, Germany
 \and deceased
\and INAF/IASF-Bologna, I-40129 Bologna, Italy
\and INAF/IASF-Roma,I-00133 Roma, Italy
\and INAF, Astronomical Obs. Brera, I-23807 Merate, Italy
\and INAF, Trieste Astronomical Observatory, I-34143 Trieste, Italy
\and INAF, Astronomical Obs. Torino, I-10025 Torino, Italy
\and University of Perugia, I-06123 Perugia, Italy
\and University of Insubria, I-22100 Como, Italy
\and INAF, IASF-Milano, I-20133 Milano, Italy
\and Abastumani Astrophysical Observatory, 383762 Abastumani, Georgia
\and now at: Scuola Normale Superiore, I-56126 Pisa, Italy
                 }

\date{Received 5 November 2009; accepted 3 March 2010}

\abstract{The blazar PG\,1553+113 is a well known TeV $\gamma$-ray emitter. In this paper we determine its spectral energy distribution through simultaneous multi-frequency data to study its emission processes. An extensive campaign was carried out between March and April 2008, where optical, X-ray, high-energy (HE) $\gamma$-ray, and very-high-energy (VHE) $\gamma$-ray data were obtained with the KVA, Abastumani, REM, \textit{Rossi}XTE/ASM, AGILE and MAGIC telescopes, respectively. We combine the data to derive the source's spectral energy distribution and interpret its double-peaked shape within the framework of a synchrotron self-Compton model.}

\keywords{
BL Lacertae objects: individual: PG\,1553+113 - Gamma rays: observation - Gamma rays: theory}


\maketitle

\section{Introduction}

The transformation  of gravitational energy in an accretion disk around a supermassive 
black hole into radiation is the widely believed underlying cause of emission in active 
galactic nuclei (AGN). Furthermore, the emission is beamed from the jet perpendicular to the disk by a 
mechanism that although not fully understood yet, most likely relates to the focusing 
properties of the rotating, fully ionized accretion disk (e.g. Blandford \& Znajek 1977). 
The viewing angle of the observer determines the observed phenomenology of AGN (Urry \& 
Padovani 1995). The AGN whose relativistic plasma jets point towards the observer are 
called blazars. The blazar class includes flat spectrum radio quasars (FSRQs) and BL Lac objects, the main difference between the two classes is in their emission lines, which are strong and quasar-like for FSRQs and weak or absent in BL Lacs.

The overall (radio-to-$\gamma$-ray) spectral energy distribution (SED) of blazars shows 
two broad non-thermal continuum peaks. For high-peaked BL Lac objects (HBLs), the 
first peak of the SED is in the UV/X-ray bands [as opposed to IR/optical  for low peaked 
BL Lac objects, LBLs], whereas the second peak is in the multi-GeV band (multi MeV for LBLs). 
The low-energy peak is thought to arise from electron synchrotron emission. The leptonic 
model sugests that the second peak forms due to inverse Compton emission. This can be due to scattering of the synchrotron photons by the non-thermal population of electrons responsible for the synchrotron radiation (synchrotron self-Compton, SSC; eg: Maraschi 1992) 
or of photons that originate outside the relativistic plasma blob (external Compton, EC; 
an external source of these \textquoteleft seed\textquoteright photons could be the accretion disk (eg: Dermer 1993) and/or the broadline region (eg: Sikora et al. 1994)). 
Blazars often show violent flux variability, which may or may not be correlated between the 
different energy bands. Strictly simultaneous observations are crucial to investigate these correlations and understand the underlying physics of blazars. 

The HBL source PG\,1553+113 was firmly detected at very high-energy $\gamma$-rays (VHE; photon energy E$>$100 GeV) by the MAGIC telescope at a significance level of 8.8\,$\sigma$ above 200 GeV, based on data from April - May 2005 and January - April 2006 (Albert et al. 2007). 
Observations with the H.E.S.S. telescope array in 2005 yielded a tentative detection in the VHE band, at the level of 4\,$\sigma$ (5.3\,$\sigma$ using a low energy threshold analysis; Aharonian et al. 2006), which was confirmed later with the combination of the 2005 and 2006 datasets (Aharonian et al. 2008).
 After the first detection of PG\,1553+113 with MAGIC, a  multi-frequency campaign on this source was conducted in July 2006 (Albert et al. 2009). The main difference between our present and the previous campaign is the use of X-ray and the high-energy (HE; photon energy E$>$100 MeV) flux.

The lack of detection of spectral lines (neither in emission nor in absorption) in the 
optical spectrum of PG\,1553+113 makes it impossible to directly measure its redshift (Falomo \& Treves 1990). However, 
an ESO-VLT spectroscopic survey of unknown-redshift BL\,Lac objects suggests $z>0.09$ (Sbarufatti et al. 2006), while the absence of host galaxy detection in HST images raises this lower limit to $z>0.25$ (Treves et al. 2007). On the other hand, the absence of a break in the VHE 
spectrum can be interpreted as suggesting $z<0.42$ (Mazin \& Goebel 2007). The absence of a Ly-$\alpha$ forest (Falomo \& Treves) in the the spectrum also constrains a lower redshift.
 The data obtained in the far-UV by the Cosmic Origins Spectrograph installed in the Hubble Space Telescope is of sufficient quality to select $\sim 40$ Ly-$\alpha$ absorbers at low redshift including a strong line at z=0.395, which constrains the resdshift of the source to be $z>0.395$ (Danforth, private comm.).

\section{Optical and near infrared (NIR) data}

\subsection{Kungliga Vetenskapsakademien observations}

The Kungliga Vetenskapsakademien (KVA, Royal Swedish Academy of Sciences) telescope
is located at the Roque de los Muchachos, in the North-Atlantic canary islands of La Palma and is operated by the Tuorla Observatory.  
The telescope is composed of a 0.6m f/15 Cassegrain devoted to polarimetry, and a 0.35m 
f/11 SCT auxiliary telescope for multicolor photometry. This telescope has been 
successfully operated remotely since the fall 2003. The KVA is used for optical (R-band) support 
observations during MAGIC observations. Typically, one measurement 
per night and per source is conducted. Photometric measurements are made in differential mode, i.e. by 
obtaining CCD images of the target and calibrated comparison stars in the same field of 
view (Fiorucci \& Tosti 1996; Fiorucci et al. 1998; Villata et al. 1998).

\subsection{Abastumani observations}
Observations at the Abastumani Observatory (Georgia, FSU) were performed with the 70 cm meniscus telescope (f/3).
This is equipped with an Apogee Ap6E CCD camera, with $390 \times 390$ pixels, and a field of view of $15 \times 15$ arcmin.
Its quantum efficiency is 40\% at 4000 \AA\ and 65\% at 6750 \AA.
The frames were acquired in the Cousins' $R$ band and were reduced with the
DAOPHOT II package\footnote{\tt http://www.star.bris.ac.uk/~mbt/daophot/}.

The source magnitude was derived from differential photometry with respect to a reference star in the same field, which lies $\sim 46$ arcsec east and $\sim 5$ arcsec south of PG 1553+113. According to the USNO 2.0 Catalogue (Monet et al. 1998), its magnitude is $R=13.2$.

\subsection{Rapid Eye Mount observations}

The Rapid Eye Mount (REM, a fast-slewing robotized infrared telescope: Covino et al. 2001) acquired photometry of PG\,1553+113 on 2008 April 18, 25 and  May 2 with all available filters (VRIJHK).  The data reduction followed standard  procedures (see e.g. Dolcini et al. 2005). The mean flux of the observation is reported in Table 3. The NIR magnitudes were calibrated against the 2MASS catalog.  For the SED reconstruction, all magnitudes were dereddened with the dust IR maps  (Schlegel et al. 1998).

\section{X-rays: {\it Rossi} X-ray Timing Explorer / All Sky Monitor observations}

The All Sky Monitor (ASM) on board the {\it Rossi} X-ray Timing Explorer ({\it R}XTE) satellite consists of three wide angle scanning shadow cameras. The cameras, mounted on a rotating drive assembly, can cover almost $70 \%$ of the sky every 1.5 hours (Levine et al. 1996). The measurements were done between 2008 March 1 and May 31. The mean measured flux of PG\,1553+113 is shown in Table 3.

\section{$\gamma$-ray data}

\subsection{HE band: AGILE observations}

The Gamma-ray Imaging Detector (GRID, 30 MeV - 30 GeV) on board the high-energy astrophysics satellite AGILE (Astro-rivelatore Gamma a Immagini LEggero; Tavani et al. 2009) observed
PG 1553+113 in three different time intervals:  2008 March 16-21,  March 25-30 and  April 10-30. The GRID data were analyzed using the AGILE standard pipeline (see Vercellone et al. 2008 for a detailed description of the AGILE data reduction), with a bin size of $0.25^{\circ} \times 0.25^{\circ}$ for E $>$ 100 MeV. Only events flagged as confirmed $\gamma$-rays and not recorded while the satellite crossed the South Atlantic Anomaly were accepted. We also rejected all events with a reconstructed direction within $10^{\circ}$ from the Earth's limb, thus reducing contamination from Earth's $\gamma$-ray albedo.
The source, observed at about 50 degrees off-axis with respect to the boresight, was not detected by the GRID at a significance level $>$ 3\, $\sigma$, and therefore the 95$\%$ confidence level upper limit was calculated. Considering that AGILE has a higher particle background at very high off-axis angles, we calculated also the upper limit selecting only photons with energies greater than 200 MeV to minimize the possible contamination at low energies. The log of the AGILE observations and the results of the analysis are reported in Table 2.
During 2008 March-April the source was outside the field of view of SuperAGILE, the hard X-ray (20-60 keV) imager onboard AGILE (Feroci et al. 2007).

\subsection{Very High Energy band: MAGIC observations}

The MAGIC Telescope (Baixeras et al. 2004, Cortina et al. 2005) is the most recent 
generation Imaging Atmospheric Cherenkov Telescope (IACT) at La Palma, Canary 
Islands, Spain (28.3$^{\circ}$N, 17.8$^{\circ}$W, 2240~m a.s.l.). Because of its
low trigger threshold of 50\,GeV (25\,GeV with a special trigger set up; Albert et al. 2008a), MAGIC is well-suited for multi-frequency 
observations together with the instruments operating in the GeV range. The 
parabolic mirror dish with a total mirror area of 236\,m$^2$ allows MAGIC to collect Cherenkov light from particle showers initiated by $\gamma$-rays or other particles in the atmosphere. 
This Cherenkov light is focused onto a multi-pixel camera composed of 577 
ultra-sensitive photomultipliers. The total field of view of the camera is 
3.5$^\circ$. The incident light pulses are converted into optical signals and 
transmitted via optical fiber to a two-level trigger system.  The selected 
events are digitized by 2~GHz Flash ADCs (Goebel et al. 2007). With a statistical analysis of the recorded light distribution and the
orientation of the shower image in the camera, the energy of the primary
particle and its incoming direction are reconstructed.  

The MAGIC observations for this campaign were carried out on 2008 March 16-18 and April 13, 28-30. The zenith angle of the data set ranges from 18 degrees to 36 degrees. 
Observations were performed in wobble mode (Fomin et al. 1994), where the object 
was observed at 0.4 degree offset from the camera center in opposite directions 
every 20 minutes. After data rejection based on the standard quality cuts and the trigger rate,  7.18 hours of total effective observation time data were selected.

An automatic analysis pipeline (Dorner at al., 2005, Bretz \& Dorner, 2008) was used to process the data, which include the muon calibration (Goebel et al. 2005) and an
absolute mispointing correction (Riegel et al. 2005). The charge distribution and arrival time information of the pulses of neighboring pixels was used to suppress the contribution from the night sky background in the shower images (Aliu et al. 2009). Three OFF regions were used to determine the background, providing a
scaling factor of 1/3 for the background calculation. The shape and orientation of the shower images were used to discriminate $\gamma$-like events from the overwhelming background. To select the $\gamma$-like events a dynamical cut in Area
(Area=$\pi\cdot$WIDTH$\cdot$LENGTH) versus SIZE (total charge contained in an image) and a cut in
$\vartheta$ (angular distance between real source position and reconstructed source position) were applied. More details on the cuts can be found in
Riegel \& Bretz (2005), and the above mentioned image parameters are described by Hillas (1985). The reconstructed $\gamma$-ray spectrum is shown in Fig.1. For the spectral reconstruction, looser cuts were applied to ensure that more than 90\% of the simulated gamma photons survived. Varying cut efficiencies between 50$\%$ and 95$\%$ over the entire energy range were applied to the data to check systematic effects of the cut efficiency on the spectral shape (shown as gray area in Fig.1). Data which were affected by calima (sand dust from the Sahara in an air
layer between 1.5 km and 5.5 km~a.s.l. causing
absorption of the Cherenkov light) were corrected following the method described in Dorner et al. (2009).


\section{ Results and Discussion}

Analyzing the MAGIC data, an excess of 415 $\gamma$-like  events, over 1835 normalized background events was found, yielding  a significance of 8.0\,$\sigma$.
The resulting differential VHE spectrum of PG\,1553+113 averaged over all observing intervals is plotted in Fig.1 (filled circles). 
It can be described by a power law 
${dN \over dE} = F_{0} \Big( {E \over 200GeV} \Big) ^{\Gamma}$ m$^{-2}$\,s$^{-1}$\,TeV$^{-1}$, where $F_{0}$ is the normalization flux at 200 GeV and $\Gamma$ is the photon index during our observation, which are both given in Table 1. The test on a possible spectral cut-off was also performed. However, fewer points of the spectrum do not favour a cut-off power law over a simple power law. The lowest point of the spectrum is at 82 GeV, mainly because of the losses of low-energy events from the cleaning and $\gamma$-selection cuts.  The values obtained during our previous observations (Albert et al. 2007) are also given in Table 1.

\begin{table}
\label{table:spec}
\begin{center}
\begin{tabular}{|p{1.in}|c|c|}
\hline
Observation period &  $F_{0}$ [ph\,TeV$^{-1}$s$^{-1}$m$^{-2}$] &  $\Gamma$   \\ \hline

March-April\,2008 & $2.0\pm0.3 \times  10^{-6}$ & $-3.4\pm0.1$ \\ \hline
March\,2008  & $1.9\pm0.4 \times 10^{-6}$& $-3.5\pm0.2$ \\ \hline
April\,2008 & $2.1\pm0.4 \times  10^{-6} $ & $-3.3\pm0.2$ \\ \hline
April-May\,2005 +  January-April\,2006 & $1.8\pm0.3 \times  10^{-6}$  & $-4.2\pm0.3$ \\ \hline

\end{tabular}
\end{center}
\caption{$F_{0}$ and $\Gamma$ during the MAGIC current observations and the previous observation. The errors are statistical only. The systematic uncertainty is estimated to be 35\% in the flux level and 0.2 in the photon index (Albert et al. 2008a).}
\end{table}

 The interaction of VHE $\gamma$-rays with the extragalactic 
background light (EBL; a recent review can be found in Mazin \& Raue 2007) leads to an attenuation of the VHE $\gamma$-ray flux via $e^+/e^-$ pair production. We computed the de-absorbed (i.e., intrinsic) fluxes with a specific \textquoteleft low 
star formation model\textquoteright of the EBL (Kneiske et al. 2004), assuming a source redshift of $z=0.3$. 
The resulting de-absorbed points are represented as empty squares in Fig.1.

\begin{figure}[h]
\label{fig:spec}
\centering
\epsfig{file=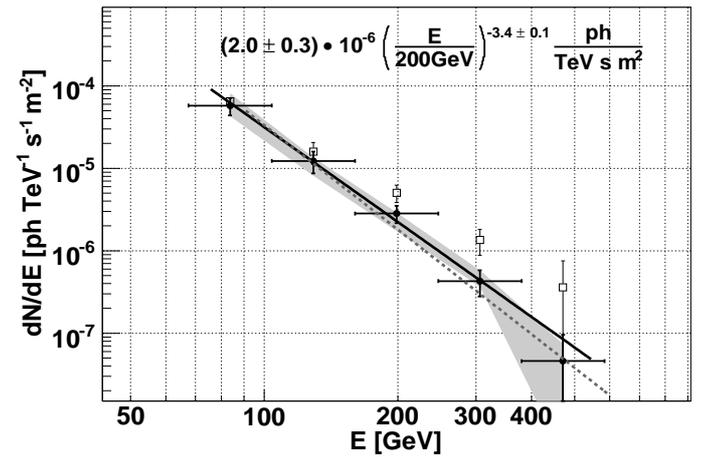,height=6cm,width=9cm,clip=}
\caption{MAGIC measured spectrum of PG\,1553+113 (filled circles). The statistical significance of the points from left to right are 2.7, 4.2, 3.4, 4.3, and 3.0 $\sigma$ respectively. The EBL-corrected points are shown as empty squares.  The spectrum obtained during our first observation is shown as a dashed line.}
\end{figure}

The HE data reduction results from AGILE are summarized in Table 2. The 2\,$\sigma$ upper limits obtained by AGILE are consistent with the average flux point observed by the $Fermi$-LAT for this source during 2008 August-Octorber (Abdo et al. 2009). The upper limit obtained in the third time interval was used for the modeling of the SED. The fluxes and corresponding effective photon frequencies of the other telescopes which contribute to this multi-frequency campaign are reported in Table 3.

\begin{table}[htdp]
\label{table:spec}
\centering
\begin{tabular}{|c|c|c|}
\hline
Time interval &  Energy &  U.L.Flux [ph\,m$^{-2}$s$^{-1}$] \\ \hline

\multirow{2}{*}{ March 16-21} &  $>$ 100\,MeV & $5.6 \times 10^{-3}$ \\
& $>$ 200\,MeV &  $3.6 \times 10^{-3}$  \\ \hline
\multirow{2}{*}{March 25-30} &  $>$ 100\,MeV & $5.5 \times 10^{-3}$ \\
& $>$ 200\,MeV &  $2.8 \times 10^{-3}$  \\ \hline
\multirow{2}{*}{April 10-30} &   $>$ 100\,MeV & $3.4 \times 10^{-3}$ \\
& $>$ 200\,MeV &  $2.1 \times 10^{-3}$  \\ \hline
\end{tabular}
\caption{2\,$\sigma$ Upper limit calculated from the AGILE data in three different time intervals.}
\end{table}

The SED of PG\,1553+113 is shown in Fig.2. The VHE and HE $\gamma$-ray 
flux points are from MAGIC and AGILE respectively. The X-ray 
point, provided by {\it R}XTE/ASM, represents the average flux between March 1 and May 31, 2008. 
The optical R-band point, provided by the KVA telescope, is the average flux obtained on 2008 March 18 and 19. The flux provided by Abastumani is the average flux of the 2008 April 1 - May 17 observations.
In addition to these  data we also used the NIR flux from REM. To assess the soundness of this addition, we checked the optical 
variability of the source during this period using Abastumani data, and found that the source was 
essentially stable (minimum and maximum values of $log(\nu F_{\nu})$ are $-10.14$ and $-10.02$ respectively). 
For a comparison of the HE flux, we included the flux points from the \textit{ Fermi} $\gamma$-ray Space Telescope (Flux, $F(E>100 MeV)= 8\pm1 \times 10^{-4} ph\,m^{-2}\,s^{-1}$ and photon index, $\Gamma=1.7\pm0.6$; Abdo et al. 2009). The average flux (15-30\,keV) obtained from the X-ray satellite $Swift$/BAT during 39 months (2004 December - 2008 February) of observation (Cusumano et al. 2010) is also included.

\begin{figure}[h]
\label{fig:ssc}
\centering
\vspace{-1cm}
\epsfig{file=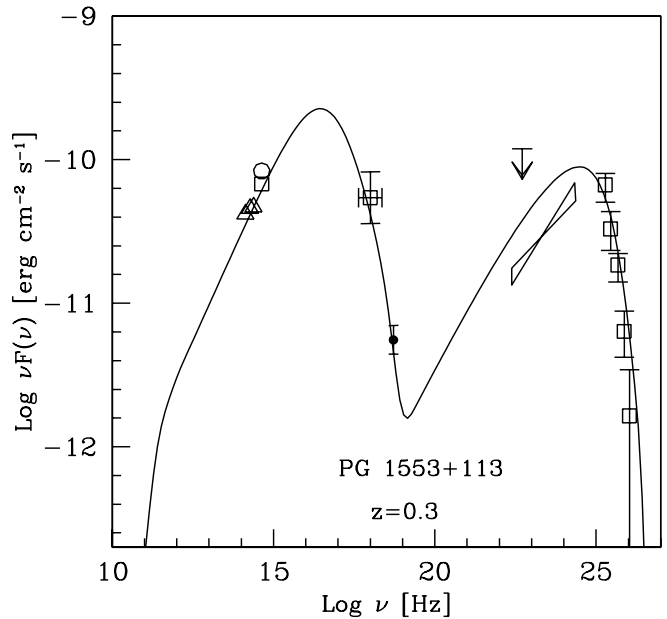,height=11cm,width=11cm,clip=}
\vspace{-1cm}
\caption{Average SED of PG\,1553+113 measured in 2008 March-April. The empty triangles denote the REM data, the open 
square represents the KVA data, the open circle denotes the Abastumani data, and the open square denotes $R$XTE/ASM data.  The arrow at HE denotes the  AGILE upper limit. The empty squares in the VHE range are the de-absorbed MAGIC data. We also show the non-simultaneous flux points from {\it Fermi} (bowtie) and $Swift$/BAT (small filled circle).}
\end{figure}

We fitted the resulting simultaneous SED  with a homogeneous one-zone SSC model (Tavecchio et al. 2001). The model assumes that  the source is
a spherical plasmon of a radius R, moving with a Doppler
factor $\delta$ towards the observer at an angle $\theta$  with respect to
the
line of sight threaded with a uniforming distributed tangled
magnetic field of the strength $B$. The injected relativistic particle population is described as a broken power-law spectrum with the normalization $K$, extending from $\gamma_{\rm 
min}$ to $\gamma_{\rm max}$ with the indices  $n_{\rm 1}$ and $n_{\rm 2}$ below and above the break Lorentz factor $\gamma_{\rm br}$. By fitting the observed flux with the model, we obtained the following parameters: 
$\gamma_{\rm min}$ = 1, 
$\gamma_{\rm br} = 3 \times 10^4$, 
$\gamma_{\rm max} = 2 \times 10^5$, 
$K=0.5 \times 10^4$ cm$^-3$, 
$n_{\rm 1}=2$, 
$n_{\rm 2}=4.7$, 
$B=0.7\,$G, 
$R=1.3 \times 10^{16}$ cm, and 
$\delta=23$. 
The optical and X-ray flux constrain on the slope of the electron energy distribution (EED), while the X-ray and VHE spectrum fix the Lorentz factors.
 
 The difference between the current SED and the previous one published in Albert et al. (2007) is due to flux variation in the X-ray and a small variation of the slope of VHE spectrum. We fitted the previous result with the Tavecchio et al. (2001) SSC model to compare the physical parameters of the SED. The difference arises from the EED, but the slopes and $\gamma_{\rm max}$ remain constant. The $\gamma_{\rm min}$ and $\gamma_{\rm br}$ of the previous observation are found to be $3\times10^3$ and  $2.7 \times 10^4$ respectively.

\begin{table}[htdp]    
\label{table:spec}
\centering
\begin{tabular}{|c|c|c|}
\hline
Instrument & log($\nu$ [Hz]) &  log($\nu$F($\nu$)) [erg cm$^{-2}$ s$^{-1}$] \\ \hline
KVA     &    14.63   &  -10.17  \\ \hline
Abastumani     &    14.63   &  -10.08  \\ \hline
\multirow{3}{*}{REM} &   14.38 & -10.33 \\
&14.27 & -10.34 \\
&14.13 & -10.38 \\ \hline
XTE & 18.03 & -10.3 \\ \hline
\end{tabular}
\caption{Effective frequencies and corresponding fluxes 
from PG\,1553+113 from KVA, Abastumani, REM and RXTE instruments obtained during this campaign.}
\end{table}
                                 
During this campaign, no significant variability of the VHE flux is found. The integral flux (E$>$200\,GeV) during these observations is $1.3 \pm 0.3 \times 10 ^{-7}$cm$^{-2}$s$^{-1}$  while during the first observations it was  $1.0 \pm 0.4 \times 10^{-7}$cm$^{-2}$s$^{-1}$. The X-ray flux\footnote{Note that the X-ray data used in Albert et al. (2007) was not taken simultaneously with VHE and optical data.} increases by about a factor of two, while the averaged X-ray flux during 39 months of $Swift$/BAT observations agrees with our SED. The optical flux during our first observation and the current observation does not show any significant variability.
The {\it Fermi} bowtie and lowest-energy MAGIC data points together with the model fit indicate a variability at HE or VHE $\gamma$-rays.

Our results suggest that the variability of PG\,1553+113 at different frequencies is time-dependent: hence only a simultaneous multi-frequency monitoring campaign over a large time span will give 
more information on the source. Relative to this it is worth mentioning that the AGILE and 
MAGIC data presented here constitute the first simultaneous broad-band $\gamma$-ray observation 
(and ensuing SED) of any blazar, though the first simultaneous detection accomplished during the multi-frequency campaign of Mkn\,421 (Donnarumma et al. 2009), and the first broad-band $\gamma$-ray spectrum was obtained from PKS\,2155-304 (Aharonian et al. 2009) by H.E.S.S. and $Fermi$.

\begin{acknowledgements} 
The MAGIC collaboration would like to thank the Instituto de Astrof{\'\i}sica de Canarias for 
the excellent working condition at the Observatorio del Roque de los Muchachos at La Palma. 
Major support from Germany's Bundesministerium f\"ur Bildung, Wissenschaft, Forschung und 
Technologie and Max-Planck-Gesellschaft, Italy's Istituto Nazionale di Fisica Nucleare (INFN) 
and Istituto Nazionale di Astrofisica (INAF), and Spain's Ministerio de Ciencia e Innovacion 
is gratefully acknowledged. The work was also supported by Switzerland's ETH Research grant 
TH34/043, Poland's Ministertwo Nauki i Szkolnictwa Wy$\dot{\rm z}$szego grant N N203 390834, and 
Germany's Young Investigator Program of the Helmholtz Gemeinschaft.
 This work was also supported by Georgian National Science
Foundation grant GNSF/ST07/4-180.
EP acknowledges support from the Italian Space Agency through grants ASI-INAF I/023/05/0 and ASI I/088/06/0.
N.M. would like to thank to C.W. Danforth for the private communication regarding the newly estimated redshift of the source.
\end{acknowledgements}

\end{document}